# Shifting $R_b$ with $A_{FB}^b$

**Darwin Chang**

*National Center for Theoretical Sciences and Department of Physics,*
*National Tsing-Hua University*
*Hsinchu 30043, Taiwan, Republic of China*

**Ernest Ma**

*Department of Physics, University of California*
*Riverside, California 92521, USA*

## Abstract

Precision measurements at the $Z$ resonance agree well with the standard model. However, there is still a hint of a discrepancy, not so much in $R_b$ by itself (which has received a great deal of attention in the past several years) but in the forward-backward asymmetry $A_{FB}^b$ together with $R_b$. The two are of course correlated. We explore the possibilty that these and other effects are due to the mixing of $b_L$ and $b_R$ with one or more heavy quarks.

Ever since the $Z$ boson was produced as a resonance at the $e^-e^+$ collider LEP at CERN, precision measurements of electroweak parameters as well as $\alpha_S$ of quantum chromodynamics (QCD) became available. Certain deviations from the predictions of the standard model have been observed in the past, notably the excess in

$$R_b \equiv \frac{\Gamma(Z \to b\bar{b})}{\Gamma(Z \to \text{hadrons})}. \tag{1}$$

This had prompted a flood of theoretical speculations regarding the possible existence of new physics[1]. At present, however, the experimental data[2] have settled down to a value of $R_b$ consistent with an excess of only $1.3\sigma$ and it is certainly not an indication of new physics by itself. On the other hand, the forward-backward asymmetry $A_{FB}^b$ is now measured to be $-2.0\sigma$ away. [This quantity used to be less accurately measured and it was always less than $\pm 1.0\sigma$ from the standard-model prediction.] If one takes seriously the two measurements together, a possible discrepancy still remains. In this paper we will explore how the mixing of $b_L$ and $b_R$ with one or more heavy quarks would explain the present data.

In the standard model, using $m_t = 175.6 \pm 5.5$ GeV and assuming $m_H = 300$ GeV, the overall best fit gives[3] $\sin^2\theta_W = 0.23152$, for which $R_b = 0.21576$ and $A_{FB}^{0,b} = 0.10308$. The experimental measurements are[2, 4]

$$R_b = 0.2170 \pm 0.0009 \quad (+1.38\sigma), \tag{2}$$

$$A_{FB}^{0,b} = 0.0984 \pm 0.0024 \quad (-1.95\sigma), \tag{3}$$

where the number of $\sigma$'s is the "pull" which is defined as the difference between measurement and fit in units of the measurement error.

Consider the couplings of the $b$ quark to the $Z$ boson:

$$\mathcal{L}_{int} = \frac{gZ^\mu}{\cos\theta_W} \left( g_L \bar{b}_L \gamma_\mu b_L + g_R \bar{b}_R \gamma_\mu b_R \right), \tag{4}$$

where the subscripts $L, R$ on $b$ refer to the left and right chiral projections $(1 \mp \gamma_5)/2$ respectively. Hence $R_b \propto g_L^2 + g_R^2$, whereas $A_{FB}^{0,b} \propto (g_L^2 - g_R^2)/(g_L^2 + g_R^2)$. ¿From the data, it is clear that a larger $g_R^2$ is desirable because that would decrease $A_{FB}^{0,b}$ and increase $R_b$. Previous attempts[5] to increase $R_b$ mostly considered increasing $g_L^2$. Two specific exceptions[6, 7] proposed to increase $g_R^2$ and we will discuss them in detail below.

In the standard model,

$$\left(g_L^2\right)_{SM} = \left(-\frac{1}{2} + \frac{1}{3}\sin^2\theta_W\right)^2 = 0.17878, \quad \left(g_R^2\right)_{SM} = \left(\frac{1}{3}\sin^2\theta_W\right)^2 = 0.00596. \quad (5)$$

In Fig. 1 we plot $R_b$ versus $A_{FB}^{0,b}$ as a function of $g_R^2(g_L^2)$ with $g_L^2(g_R^2)$ fixed at its standard-model value. The experimental range is also displayed. For $g_L^2$ fixed at $(g_L^2)_{SM}$, the best fit is

$$g_R^2 = 0.00736 \quad (6)$$

for which $R_b = 0.2174$ and $A_{FB}^{0,b} = 0.1015$ are obtained. If we let both $g_L^2$ and $g_R^2$ be free parameters, then we get the central values of $R_b$ and $A_{FB}^{0,b}$ with

$$g_L^2 = 0.17586, \quad g_R^2 = 0.00994. \quad (7)$$

We now discuss how the above two cases, *i.e.* Eqs. (6) and (7), may be obtained. In Ref. [6], a vector doublet of quarks with the conventional charges, *i.e.* 2/3 and $-1/3$, is added. We call this Model (A) with $(Q_1, Q_2)_{L,R} \sim (3, 2, 1/6)$ under $SU(3)_C \times SU(2)_L \times U(1)_Y$. Since $(Q_1, Q_2)_L$ transforms in the same way as the known quark doublets, we define it precisely as the one that forms an invariant mass with $(Q_1, Q_2)_R$. Hence the mass matrix linking $(\bar{b}_L, \bar{Q}_{2L})$ with $(b_R, Q_{2R})$ is given by

$$\mathcal{M}_{b,Q_2} = \begin{pmatrix} m_b & 0 \\ m_{Qb} & M \end{pmatrix}, \quad (8)$$

which shows that $b_R$-$Q_{2R}$ mixing is dominant, and that $b_L$-$Q_{2L}$ mixing is suppressed by $m_b/m_Q$ and is thus negligible[1]. We now have

$$g_R^2 = \left[\frac{1}{3}\sin^2\theta_W \cos^2\theta_2 + \left(-\frac{1}{2} + \frac{1}{3}\sin^2\theta_W\right)\sin^2\theta_2\right]^2 = \left[\frac{1}{3}\sin^2\theta_W - \frac{1}{2}\sin^2\theta_2\right]^2. \quad (9)$$

In order to increase $g_R^2$ from its standard-model value, it is clear that $\sin^2 \theta_2$ must be greater than $(4/3) \sin^2 \theta_W$. Hence a rather large mixing with $Q_2$ is required in this model. Numerically, to obtain Eq. (6), we need

$$\sin^2 \theta_2 = 0.3260. \tag{10}$$

In Ref. [7], a vector doublet of quarks with the unconventional charges $-1/3$ and $-4/3$ is added. We call this Model (B) with $(Q_3, Q_4)_{L,R} \sim (3, 2, -5/6)$. The $b$-$Q_3$ mass matrix is of the same form as Eq. (8) because there cannot be a $\bar{b}_L Q_{3R}$ term for lack of a Higgs triplet. In this case,

$$g_R^2 = \left[\frac{1}{2} \sin^2 \theta_W \cos^2 \theta_3 + \left(\frac{1}{2} + \frac{1}{3} \sin^2 \theta_W\right) \sin^2 \theta_3\right]^2 = \left[\frac{1}{3} \sin^2 \theta_W + \frac{1}{2} \sin^2 \theta_3\right]^2. \tag{11}$$

Now we need only a small mixing to obtain Eq. (6), namely

$$\sin^2 \theta_3 = 0.0173. \tag{12}$$

For comparison against the above two vectorial models, we consider also the addition of one mirror family of heavy fermions. The heavy quarks here are right-handed doublets and left-handed singlets. We call this Model (C) with $(Q_5, Q_6)_R \sim (3, 2, 1/6)$, $Q_{5L} \sim (3, 1, 2/3)$, and $Q_{6L} \sim (3, 1, -1/3)$. The $b$-$Q_6$ mass matrix is then

$$\mathcal{M}_{b,Q_6} = \begin{pmatrix} m_b & m_{bQ} \\ m_{Qb} & m_Q \end{pmatrix}, \tag{13}$$

which allows both $b_R$-$Q_{6R}$ and $b_L$-$Q_{6L}$ mixings, so that Eq. (7) may be satisfied. However, it is a somewhat unnatural solution because $m_b$ and $m_Q$ come from the vacuum expectation value of the Higgs doublet, whereas $m_{bQ}$ and $m_{Qb}$ are invariant mass terms. It is thus difficult to understand why the latter two masses are not much greater. Using

$$g_L^2 = \left[\left(-\frac{1}{2} + \frac{1}{3} \sin^2 \theta_W\right) \cos^2 \theta_{6L} + \frac{1}{3} \sin^2 \theta_W \sin^2 \theta_{6L}\right]^2 = \left[-\frac{1}{2} + \frac{1}{3} \sin^2 \theta_W + \frac{1}{2} \sin^2 \theta_{6L}\right]^2, \tag{14}$$

and Eq. (9) with $\theta_2$ replaced by $\theta_{6R}$ to fit Eq. (7), we find

$$\sin^2 \theta_{6L} = 0.0189, \quad \sin^2 \theta_{6R} = 0.3537. \tag{15}$$

In Model (A) and Model (C), large mixing of $b_R$ with a heavy quark is required, as shown in Eqs. (10) and (15) respectively. This has important implications on the electroweak oblique parameters $S, T, U$ or $\epsilon_1, \epsilon_2, \epsilon_3$. In Model (A), assuming that $Q_1$ does not mix with $t$, $c$, or $u$, we have the following physical doublets:

$$\begin{pmatrix} Q_1 \\ Q_2 \end{pmatrix}_L, \quad \begin{pmatrix} Q_1 \\ Q_2 \cos \theta_2 - b \sin \theta_2 \end{pmatrix}_R, \tag{16}$$

which would contribute to $T$ or $\epsilon_1$. In the above, the masses of $Q_1$ and $Q_2$ are related by $m_1 = m_2 \cos \theta_2$, assuming that $m_b << m_1, m_2$. Let $x \equiv \sin^2 \theta_2$, then we find

$$\Delta \epsilon_1 = \frac{3\alpha}{16\pi \sin^2 \theta_W} \frac{m_2^2}{M_W^2} F(x), \tag{17}$$

where

$$F(x) = -2(1-x)(2-x)\left[1 + \frac{\ln(1-x)}{x}\right] - 2x + 3x^2. \tag{18}$$

Note that $F(0) = 0$ and $F(1) = 1$ as expected. Also, $F(x) > 0$ for $0 < x < 1$. Taking $x = 0.3260$ as in Eq. (10), we get $F_1(x) = 0.141$. Let us choose $m_1 = 200$ GeV so that the decay $Q_1 \rightarrow b + W$ would not be a significant contribution to the top signal at the Tevatron. In that case, $m_2 = 244$ GeV and $\Delta \epsilon_1 = 2.6 \times 10^{-3}$ which would take this model far away[3] from the data. Since our purpose is to find out if mixing with heavy quarks would improve the overall agreement with data, this numerical result tells us that Model (A) as it stands is not the answer.

In Model (C), the physical doublets are

$$\begin{pmatrix} Q_5 \\ Q_6 \cos \theta_{6R} - b \sin \theta_{6R} \end{pmatrix}_R, \quad \begin{pmatrix} t \\ b \cos \theta_{6L} - Q_6 \sin \theta_{6L} \end{pmatrix}_L, \tag{19}$$

but since $\theta_{6L}$ is small, it can be neglected, and $m_5$ is unrelated to $m_6$. We now find

$$\Delta\epsilon_1 = \frac{3\alpha}{16\pi\sin^2\theta_W}\frac{1}{M_W^2}\left[m_5^2 + m_6^2\cos^4\theta_{6R} - \frac{2m_5^2 m_6^2\cos^2\theta_{6R}}{m_5^2 - m_6^2}\ln\frac{m_5^2}{m_6^2}\right]. \quad (20)$$

Hence we can fine-tune $m_5^2/m_6^2$ to make $\Delta\epsilon_1$ small. For example, if we let $m_5 = 200$ GeV, then the above expression is minimized with $m_6 = 273$ GeV for which $\Delta\epsilon_1 = 0.52 \times 10^{-3}$. This much smaller shift is acceptable. On the other hand, unlike Models (A) and (B) where the heavy quarks are doublets in both left and right chiralities, Model (C) has $Q_{5L}$ and $Q_{6L}$ as singlets, hence the shift in $S$ or $\epsilon_3$ becomes nonnegligible. Let $x \equiv \sin^2\theta_{6R}$ and assume $M_Z << m_5, m_6$, then we find

$$\Delta\epsilon_3 = \frac{\alpha}{24\pi\sin^2\theta_W}\left[3 - 8x + 5x^2 - \ln\frac{m_5^2}{m_6^2} - x(2 - 3x)\ln\frac{m_b^2}{m_6^2}\right] = 1.8 \times 10^{-3} \quad (21)$$

for $x = 0.3537$, $m_5 = 200$ GeV, and $m_6 = 273$ GeV. This shift would already take this model far away[3] from the data, not to mention that there is also the leptonic contribution of $0.44 \times 10^{-3}$. Hence Model (C) is also not the answer.

Let us go back to Model (A) and try to reduce $\Delta\epsilon_1$ of Eq. (17) by allowing $Q_1$ to mix with $t$. In that case, we have

$$\begin{pmatrix} Q_1\cos\theta_{1R} - t\sin\theta_{1R} \\ Q_2\cos\theta_{2R} - b\sin\theta_{2R} \end{pmatrix}_R, \begin{pmatrix} Q_1\cos\theta_{1L} - t\sin\theta_{1L} \\ Q_2 \end{pmatrix}_L, \begin{pmatrix} t\cos\theta_{1L} + Q_1\sin\theta_{1L} \\ b \end{pmatrix}_L, \quad (22)$$

where the masses of $Q_1$, $Q_2$, and $t$ are related by

$$m_1 = M(\cos\theta_{1L}/\cos\theta_{1R}), \quad m_2 = M/\cos\theta_{2R}, \quad m_t = M(\sin\theta_{1L}/\sin\theta_{1R}), \quad (23)$$

where $M$ is defined as in Eq. (8). After a straightforward calculation, we find

$$\begin{aligned}\Delta\epsilon_1 &= \frac{3\alpha}{8\pi\sin^2\theta_W}\frac{1}{M_W^2}\left[-4M^2 + \frac{1}{2}m_1^2(1 + c_{1R}^4) + \frac{1}{2}m_2^2(1 + c_{2R}^4) + \frac{1}{2}m_t^2 s_{1R}^4 \right. \\ &\quad \left. + A_1 m_1^2 \ln m_1^2 + A_2 m_2^2 \ln m_2^2 + A_t m_t^2 \ln m_t^2\right],\end{aligned} \quad (24)$$

where $c_{1R} \equiv \cos\theta_{1R}$, $s_{1R} \equiv \sin\theta_{1R}$, etc., and

$$A_1 = -c_{1R}^2 s_{2R}^2 - s_{1L}^2 + s_{1R}^4 + (4c_{1R}^2 - c_{1R}^2 c_{2R}^2 - c_{1L}^2)\frac{m_1^2}{m_1^2 - m_2^2} + (c_{1R}^2 s_{1R}^2)\frac{m_1^2}{m_1^2 - m_t^2}, \quad (25)$$

$$A_2 = s_{2R}^2 + (-4c_{1L}^2 c_{2R}^2 + c_{1R}^2 c_{2R}^2 + c_{1L}^2)\frac{m_2^2}{m_1^2 - m_2^2} + (4s_{1L}^2 c_{2R}^2 - s_{1R}^2 c_{2R}^2 - s_{1L}^2)\frac{m_2^2}{m_2^2 - m_t^2}, \quad (26)$$

$$A_t = c_{1R}^4 - s_{1R}^2 s_{2R}^2 - c_{1L}^2 - (c_{1R}^2 s_{1R}^2)\frac{m_t^2}{m_1^2 - m_t^2} + (-4s_{1R}^2 + s_{1R}^2 c_{2R}^2 + s_{1L}^2)\frac{m_t^2}{m_2^2 - m_t^2}. \quad (27)$$

Note that $A_1 m_1^2 + A_2 m_2^2 + A_t m_t^2 = 0$ as expected. We now let $m_1 = 200$ GeV and using $m_t = 175.6$ GeV, we have

$$\frac{s_{1L}^2}{c_{1L}^2} = \left(\frac{175.6}{200}\right)^2 \frac{s_{1R}^2}{c_{1R}^2}. \quad (28)$$

For a given value of $c_{1R}^2$, we then fix $c_{1L}^2$ and hence $M$ $(= m_1 c_{1R}/c_{1L})$ as well as $m_2$ $(= M/c_{2R})$, assuming of course that $c_{2R}^2 = 0.6740$ from Eq. (10). We vary $c_{1R}^2$ and compute the right-hand side of Eq. (24) numerically. We find that it is in fact a monotonically increasing function of decreasing $c_{1R}^2$. Hence the value obtained earlier for $\Delta\epsilon_1$ assuming no $Q_1$ mixing (i.e. $c_{1R}^2 = c_{1L}^2 = 1$), which was already too far away from the experimental data, cannot be reduced, and Model (A) is not saved by additional mixings.

Now that both Models (A) and (C) are eliminated by the precision electroweak measurements, we focus our remaining discussion on Model (B). The additional heavy quarks $Q_3$ and $Q_4$ have charges $-1/3$ and $-4/3$ respectively. The physical doublets are

$$\begin{pmatrix} Q_3 \\ Q_4 \end{pmatrix}_L, \quad \begin{pmatrix} Q_3 \cos\theta_3 - b\sin\theta_3 \\ Q_4 \end{pmatrix}_R, \quad (29)$$

whereas

$$b_R \cos\theta_3 + Q_{3R} \sin\theta_3 \quad (30)$$

is a singlet. Hence $Q_4$ decays into $b + W^-$ and would have been observed in the top-quark search at the Tevatron if its mass is below 200 GeV or so. In fact, the Tevatron top-quark events cannot tell $b$ from $\bar{b}$, hence it is even conceivable that $Q_4$ was actually discovered instead of $t$. However, $Q_3$ would also have been produced, since $m_3 = m_4/\cos\theta_3 \simeq 1.01 m_4$, and it would decay into either $b + Z$ or $b + H$. The nonobservation of the $b + Z$ mode at CDF requires the $b + H$ mode to dominate. However, since LEP data already require the Higgs scalar mass to be greater than about 90 GeV, the $b + H$ mode cannot dominate and this exotic possibility is ruled out. We conclude that $Q_3$ and $Q_4$ are hitherto undiscovered and must be heavier than about 200 GeV. Note that this is beyond the reach of LEP for producing $\bar{b}Q_3 + b\bar{Q}_3$.

We have so far assumed that $Q_3$ mixes only with $b$, but of course it could also mix with $s$ and $d$. In that case, the state $b$ in Eqs. (29) and (30) should be considered as a linear combination of $b$, $s$, and $d$, but dominated by $b$. As a result, there would be flavor-nondiagonal couplings of the $Z$ to $d\bar{s}+s\bar{d}$, $d\bar{b}+b\bar{d}$, and $s\bar{b}+b\bar{s}$, and in addition, the charged-current mixing matrix mediated by $W$ would lose its unitarity. These couplings are presumably very small, but they could affect the standard-model phenomenology regarding the $K^0 - \bar{K}^0$, $B^0 - \bar{B}^0$, and $B_s^0 - \bar{B}_s^0$ systems. Furthermore, the indirect effect of flavor-nondiagonal neutral currents in the right-handed sector could result in the failure of the Standard Model to describe all data, especially the precision measurements to be obtained with the upcoming $B$ factories, at KEK and at SLAC. For example, the exotic contributions to the decay $B \to X_s \gamma$ in vector quark models [including our Model (B)] have been analyzed recently[8]. The dominant extra contribution is from the violation of unitarity in the charged-current mixing matrix. It gives rise to nontrivial constraints on the off-diagonal mass matrix elements $b$-$Q_3$ and $s$-$Q_3$. Note that whereas the Standard Model predicts a branching fraction of $b \to s\gamma$, including the next-to-leading-order correction, which is still allowed by the experimental data, future reduction in the experimental error with the same central value may be a potential signal for new vector quarks.

In conclusion, there may still be a hint of new physics in the current precision measurements of $R_b$ and $A_{FB}^b$. If it is due to the mixing of $b$ with heavy quarks, the only viable model[7] is to add a heavy vector doublet of quarks with the unconventional charges $-1/3$ and $-4/3$. Two other models are eliminated because they require large mixings, which in turn generate large shifts in $\epsilon_1$ and $\epsilon_3$, and are thus in disagreement with present precision data.

## ACKNOWLEDGEMENT

This work was supported in part by the U. S. Department of Energy under Grant No. DE-FG03-94ER40837 and by a grant from the National Science Council of R.O.C. We thank G. P. Yeh and C.-H. V. Chang for discussions. DC thanks the Physics Department, U.C. Riverside for hospitality during a visit when this work was initiated.

# References

[1] For a review, see for example P. Bamert, C. P. Burgess, J. M. Cline, D. London, and E. Nardi, Phys. Rev. **D54**, 4275 (1996).

[2] The LEP Collaborations *et al.*, CERN-PPE/97-154 (Dec 97).

[3] G. Altarelli, R. Barbieri, and F. Caravaglios, hep-ph/9712368 (Dec 97).

[4] D. Ward, in Proc. of the EPS High Energy Physics Meeting, Jerusalem, Israel (Aug 97).

[5] See for example E. Ma, Phys. Rev. **D53**, R2276 (1996) and many other references found in Ref. [1].

[6] T. Yoshikawa, Prog. Theor. Phys. **96**, 269 (1996).

[7] C.-H. V. Chang, D. Chang, and W.-Y. Keung, Phys. Rev. **D54**, 7051 (1996).

[8] C.-H. V. Chang, D. Chang, and W.-Y. Keung, "Vector Quark Model and B Meson Radiative Decay", unpublished.

FIGURE CAPTION

Fig. 1. Plot of $R_b$ versus $A_{FB}^{0,b}$. The solid lines are the experimental ranges. The dashed (dotted) line is obtained by varying $g_L^2$ ($g_R^2$) holding $g_R^2$ ($g_L^2$) fixed in the Standard Model.

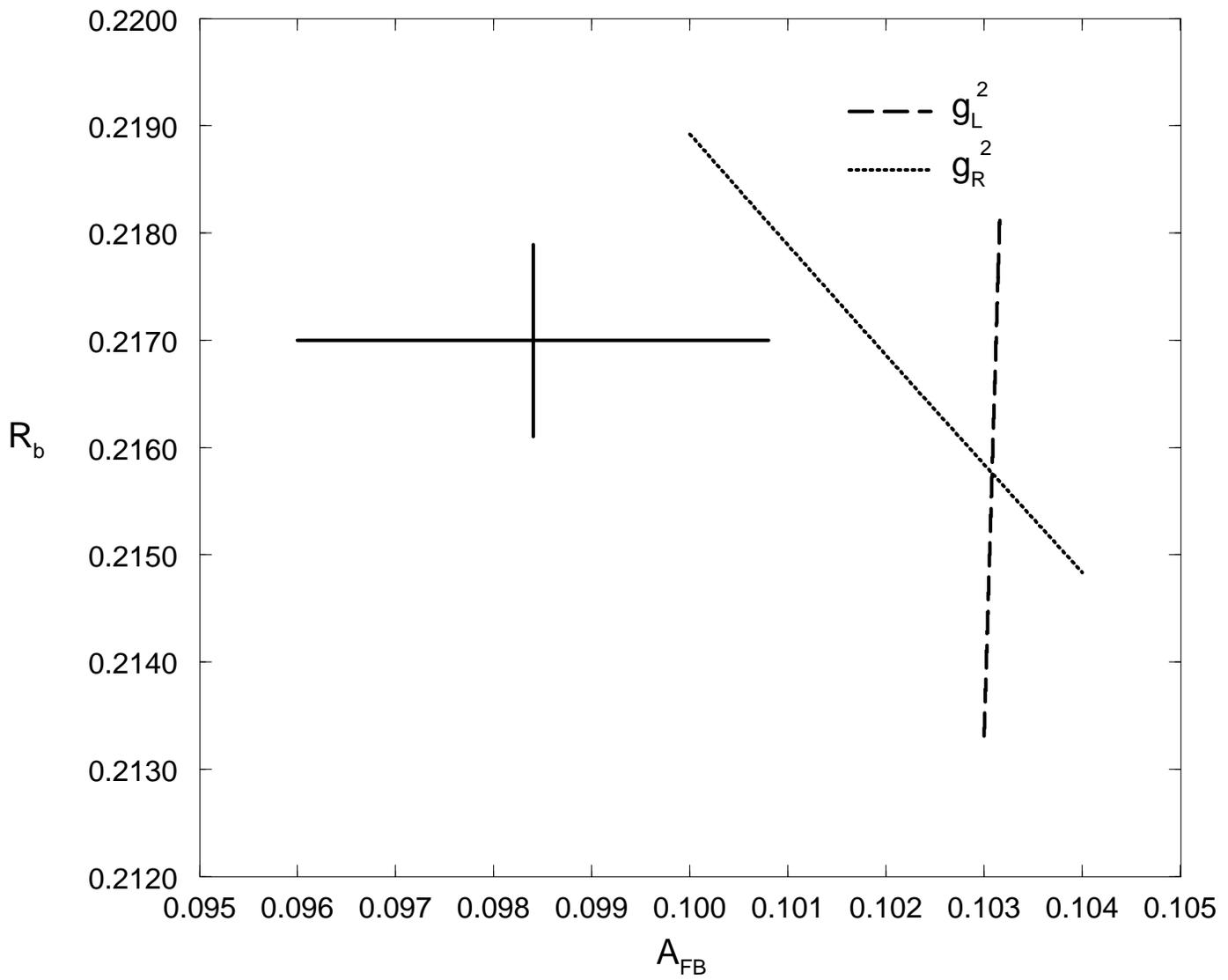